\documentclass[11pt]{article}

\usepackage[final]{acl}

\usepackage{times}
\usepackage{latexsym}

\usepackage[T1]{fontenc}
\usepackage[utf8]{inputenc}

\usepackage{microtype}

\usepackage{inconsolata}

\usepackage{graphicx}

\usepackage{cite}
\usepackage{amsmath,amssymb,amsfonts}
\usepackage{graphicx}
\usepackage{textcomp}
\usepackage{xcolor}
\usepackage[utf8]{inputenc} % allow utf-8 input
\usepackage[T1]{fontenc}    % use 8-bit T1 fonts
\usepackage{hyperref}
\usepackage{url}            % simple URL typesetting
\usepackage{booktabs}       % professional-quality tables
\usepackage{nicefrac}       % compact symbols for 1/2, etc.
\usepackage{microtype}      % microtypography
\usepackage{wasysym}

\def\equationautorefname#1#2\null{(#2\null)}

\usepackage{comment}
\usepackage{algorithmicx,algorithm}
\usepackage{algpseudocode}
\usepackage{mathtools}
\usepackage{threeparttable}
\usepackage{multirow}
\usepackage{colortbl} 
\usepackage{natbib}
\usepackage{subcaption}
\usepackage{float}

\title{ZipVoice-Dialog: Non-Autoregressive Spoken Dialogue Generation \\ with Flow Matching}

\author{Han Zhu, Wei Kang, Liyong Guo, Zengwei Yao, Fangjun Kuang \\
\textbf{Weiji Zhuang, Zhaoqing Li, Zhifeng Han, Dong Zhang, Xin Zhang} \\
\textbf{Xingchen Song, Lingxuan Ye, Long Lin, Daniel Povey} \\
  Xiaomi Corp., Beijing, China \\
  \texttt{\{zhuhan3,dpovey\}@xiaomi.com}}

\begin{document}
\maketitle
\begin{abstract}
Generating spoken dialogue is inherently more complex than monologue text-to-speech (TTS), as it demands both realistic turn-taking and the maintenance of distinct speaker timbres. While existing autoregressive (AR) models have made progress, they often suffer from high inference latency and stability issues. To overcome these limitations, we propose ZipVoice-Dialog, a non-autoregressive (NAR) zero-shot spoken dialogue generation model based on flow-matching. Observing that applying vanilla flow-matching to dialogue generation leads to poor speech intelligibility and turn-taking precision, we introduce two simple yet effective methods to adapt flow-matching architectures for dialogue generation: (1) a curriculum learning strategy to ensure robust speech-text alignment, and (2) speaker-turn embeddings to govern precise speaker turn-taking. Additionally, we introduce dedicated strategies to support stereo dialogue generation.
Recognizing the lack of training datasets in this field, we curate and release OpenDialog, the first large-scale (6.8k hours) open-source spoken dialogue dataset derived from in-the-wild speech data. Moreover, for fair and rigorous evaluations, we established a benchmark to comprehensively evaluate dialogue generation models. Experiments demonstrate the effectiveness of the proposed methods and dataset, showing that ZipVoice-Dialog achieves superior performance in inference speed, intelligibility, speaker turn-taking accuracy, and speaker similarity. Our code, model checkpoints, and the OpenDialog dataset are publicly available\footnote{https://github.com/k2-fsa/ZipVoice}.
\end{abstract}

\section{Introduction}
Recent advancements in text-to-speech (TTS) have enabled the generation of highly natural monologue speech~\citep{anastassiou2024seed,shennaturalspeech,du2024cosyvoice2,wang2023neural,le2023voicebox,eskimez2024e2,chen2024f5,zhu2025zipvoice,wang2024maskgct,wang2025spark,deng2025indextts,guo2024fireredtts}. However, synthesizing spontaneous spoken dialogues involving multiple speakers remains a significant challenge.  This difficulty primarily stems from two aspects. Firstly, spoken dialogues require accurate and natural speaker turn-taking~\citep{nguyen2023generative}, which, despite being intuitive for humans, is non-trivial for TTS models to capture. Secondly, spoken dialogues inherently involve multiple speakers with distinct timbres, creating acoustic complexities that hinder the learning of speech-text alignment.

Current state-of-the-art methods predominantly rely on autoregressive TTS architectures for dialog generation~\citep{ju2025mooncast,zhang2024covomix,darefsky2024parakeet,DiaRepo}. While effective, these models inherently suffer from high inference latency due to their sequential nature. Furthermore, AR models are prone to exposure bias, often leading to robustness issues such as word repetition or skipping~\citep{song2025ella,yang2025pseudo}.

To address these limitations, we propose ZipVoice-Dialog, a flow-matching-based non-autoregressive (NAR) zero-shot model designed for efficient and stable spoken dialogue generation.
Although flow-matching has shown promise in monologue TTS, we observe that applying vanilla flow-matching architectures directly to dialogue tasks results in unintelligible speech with unstable turn-taking.
To bridge this gap, we propose two simple yet effective designs to make flow-matching-based architectures suitable for dialogue generation.

Firstly, to mitigate the difficulty of learning speech-text alignments across multiple timbres, we propose a staged training strategy. By first establishing a foundation on monologue data and subsequently fine-tuning on dialogue speech, ZipVoice-Dialog inherits robust speech-text alignment while mastering conversational characteristics.

Secondly, accurate speaker turn-taking, i.e., assigning the correct speaker voice to each part of the dialogue, is an essential characteristic of natural spoken dialogues. To achieve this, we propose the incorporation of two learnable speaker-turn embeddings into the text conditioning. These embeddings provide explicit and detailed speaker turn-taking cues to the model. Despite its simplicity, this method significantly improves speaker turn-taking accuracy.

Furthermore, we extend ZipVoice-Dialog’s capabilities to stereo dialogue generation by employing a weight initialization strategy,  a single-channel dialogue regularization method, and a speaker-exclusive loss.

A critical bottleneck in this field is the scarcity of large-scale, high-quality dialogue data.
We address this by curating and releasing OpenDialog, a 6.8k-hour open-source spoken dialogue dataset derived from diverse in-the-wild sources.

To thoroughly evaluate spoken dialogue generation models, we designed a comprehensive evaluation benchmark. This benchmark incorporates real-world evaluation data and various reproducible objective metrics to standardize the assessment of dialogue quality.

Experimental results demonstrate that ZipVoice-Dialog achieves state-of-the-art performance, outperforming existing models like MoonCast~\citep{ju2025mooncast} and Dia~\citep{DiaRepo} in terms of inference speed, intelligibility, speaker turn-taking accuracy, and speaker similarity

Our main contributions are summarized as follows:

\begin{itemize}
    \item We propose ZipVoice-Dialog, a NAR flow-matching-based model that achieves fast, stable, and high-quality zero-shot  dialogue generation, overcoming the robustness issues and speed limitations of AR baselines.

    \item We introduce two simple yet effective strategies to enable dialogue generation with flow-matching architecture: (1) a curriculum learning strategy to guarantee stable speech-text alignment, and (2) learnable speaker-turn embeddings to ensure precise turn-taking.

    \item We release OpenDialog, the first large-scale (6.8k hours) open-source spoken dialogue dataset.

    \item We establish a comprehensive benchmark with real-world dialogue data and various reproducible objective metrics.
\end{itemize}

\section{Related Works}

\subsection{Non-Autoregressive Monologue TTS}

In the field of monologue TTS, where one utterance features a single speaker, non-autoregressive (NAR) architectures~\citep{le2023voicebox,eskimez2024e2,chen2024f5,zhu2025zipvoice,wang2024maskgct,ren2019fastspeech} have gained considerable traction. Compared to their AR counterparts, NAR models offer superior inference speed and enhanced stability through parallel generation.
Among these, flow matching~\citep{lipman2023flow} has emerged as a particularly effective framework, achieving high-quality synthesis with simpler training objectives and fewer sampling steps than traditional diffusion models~\citep{song2020score}. Consequently, flow matching has become the backbone of various state-of-the-art (SOTA) NAR TTS systems~\citep{le2023voicebox,eskimez2024e2,chen2024f5,zhu2025zipvoice}.
Our work builds upon this flow-matching-based paradigm to extend NAR capabilities to multi-speaker contexts.

\subsection{Spoken Dialogue Generation}

Unlike monologue TTS, spoken dialogue generation involves synthesizing multi-turn interactions among multiple speakers.
To achieve natural turn-taking, early research focused on "textless" spoken language models~\citep{nguyen2023generative,meng2024parrot,fu2024investigating}. While these models excel at generating naturalistic turn-taking, they lack direct semantic control via text input.

In contrast, conversational TTS models~\citep{guo2021conversational,cong2021controllable,liu2024generative,xue2023m} generate speech one utterance at a time conditioned on linguistic or acoustic dialogue history. This approach, however, fails to capture the global structure of a dialogue, potentially compromising overall naturalness.

To bridge this gap, recent studies have explored text-conditioned dialogue generation.
Some extend dialogue generative spoken language models (dGSLM)~\citep{nguyen2023generative,mitsui2023towards,lu2025slide} to incorporate text conditions, while others adapt AR-based monologue architectures for dialogue tasks~\citep{ju2025mooncast,zhang2024covomix,darefsky2024parakeet,DiaRepo}. Despite their progress, these AR models suffer from two primary drawbacks: 1) computational inefficiency due to sequential sampling, and 2) instability, where unidirectional modeling and exposure bias lead to robustness issues such as word skipping or repetition~\citep{song2025ella,yang2025pseudo}.

While NAR architectures have gained considerable popularity in monologue TTS, their applicability to spoken dialogue generation remains underexplored. As demonstrated by our preliminary experiments (see \autoref{tab:curriculum}), directly applying the flow-matching based TTS architecture for dialogue scenarios results in unintelligible speech. This underscores the challenge of designing NAR architectures specifically for the complexities of dialogue.  A concurrent work~\citep{zhang2025covomix2} also explored this direction by generating spoken dialogue given pre-defined timestamps of each speaker turn. Different from their work, we explored an end-to-end NAR solution that does not rely on pre-defined timestamps, thus being simpler in training (does not rely data with speaker-turn timestamps) and inference (does not rely on additional timestamps prediction models).

\section{Preliminary: Flow-Matching TTS}

ZipVoice-Dialog is based on the flow-matching-based TTS paradigm. In this section, we briefly review this architecture, specifically focusing on ZipVoice~\citep{zhu2025zipvoice}, a state-of-the-art monologue TTS model that serves as our foundation.

In terms of model architecture, ZipVoice comprises a text encoder and a vector field estimator, both of which utilize the Zipformer~\citep{yaozipformer} as their backbone. For waveform synthesis, a pre-trained Vocos~\citep{siuzdak2024vocos} vocoder is employed to convert the generated acoustic features into high-fidelity speech.

ZipVoice is trained with a conditional flow matching (CFM) objective~\citep{le2023voicebox}.
And speech infilling task~\citep{le2023voicebox} is adopted to enable zero-shot generation ability. Specifically, given an tokenized text sequence $y=(y_1, y_2, \ldots, y_N)$, the text encoder extracts text features $\hat{y} \in \mathbb{R}^{F \times N}$. To align with the temporal dimension of the speech features $x_1 \in \mathbb{R}^{D \times T}$, $\hat{y}$ is expanded via average upsampling to form the text condition $z \in \mathbb{R}^{F \times T}$, assuming a uniform duration for each token.

The model learns to reconstruct masked segments of the speech, defined by a binary mask $m \in \{0, 1\}^{D \times T}$ (where 1 indicates masked positions). The vector field estimator $v_t$ takes three concatenated inputs: the unmasked speech context $(1-m) \odot x_1$, the text condition $z$, and the interpolated noisy features $x_t = (1-t)x_0 + tx_1$, where $t \in [0, 1]$ is the timestep and $x_0 \sim \mathcal{N}(0, I)$. The training objective is to minimize the following CFM loss:

\begin{equation}
\label{equ:cfm_tts}
\begin{aligned}
L_{\text{CFM-TTS}} &= \mathbb{E}_{t, q(x_1), p_0(x_0)} \\
&\quad \times \bigg\lVert \bigg( v_t\big( x_t, z, (1-m)\odot x_1; \theta\big) \\
&\quad - (x_1-x_0) \bigg) \odot m \bigg\rVert^2
\end{aligned}
\end{equation}
Note that the loss is computed only over the masked regions $m$ to focus the model on generating the missing speech segments.

During inference, the total duration is estimated based on the token length ratio between the target text and the prompt. ZipVoice generates speech features using an Euler ODE solver, along with a time-dependent classifier-free guidance (CFG)~\citep{ho2021classifierfree} strategy.

\section{Proposed Method: ZipVoice-Dialog}

\begin{figure*}[t!]
	\centering
        \begin{subfigure}[b]{0.99\columnwidth}
	\includegraphics[width=\columnwidth]{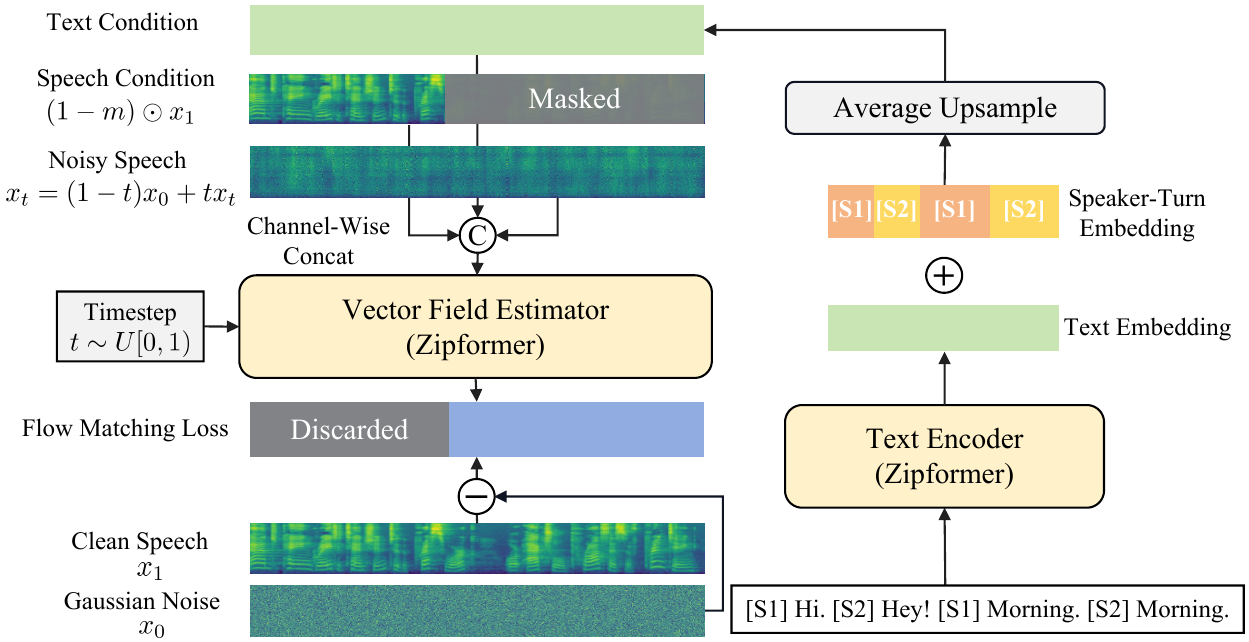}
    \end{subfigure}
    \hfill 
    \begin{subfigure}[b]{0.96\columnwidth}
	\includegraphics[width=\columnwidth]{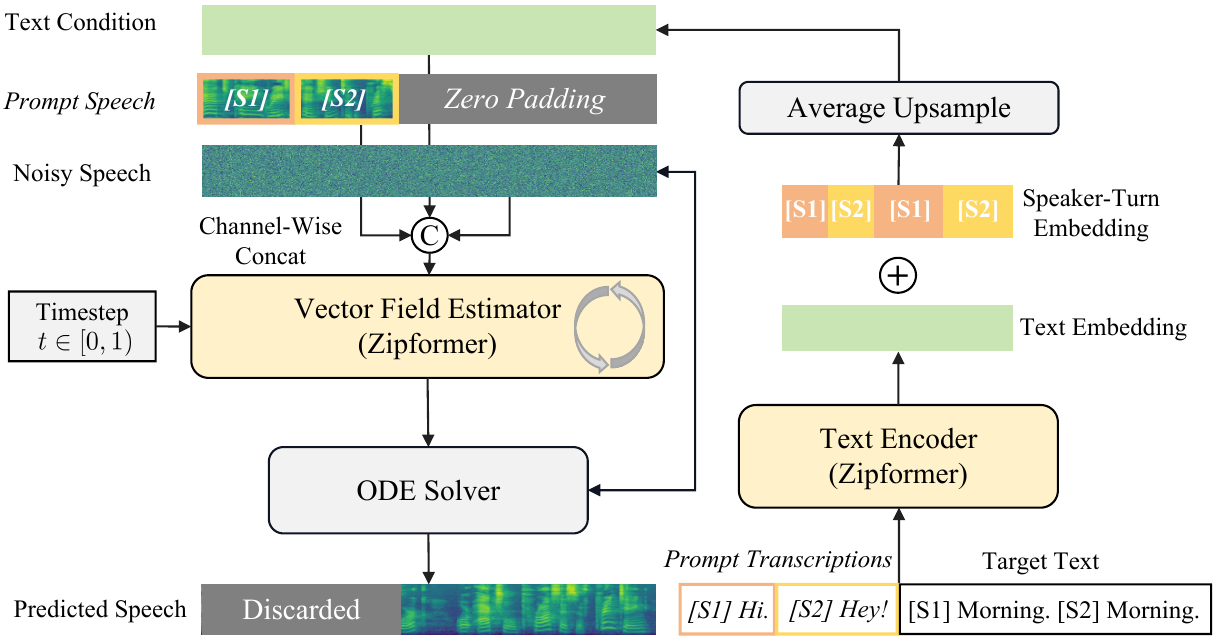}
    \end{subfigure}
	\caption{Illustration of ZipVoice-Dialog training (left) and inference (right).} 
	\label{fig:architecture}
\end{figure*} 

This section details the ZipVoice-Dialog architecture. Its training and inference pipelines are illustrated in Figure~\ref{fig:architecture}. Below, we discuss the key designs that differentiate ZipVoice-Dialog from conventional monologue flow-matching-based TTS models.

\subsection{Monologue-to-Dialogue Curriculum Learning}

Our preliminary experiments show that directly training a flow-matching model on dialogue data leads to alignment collapse, yielding unintelligible speech.
This is primarily due to the difficulty of learning speech-text alignments in conversations involving two distinct speakers. To mitigate this, we employ a two-stage curriculum learning strategy that first establishes alignment capabilities and then captures conversational dynamics:

\textbf{Stage 1: Monologue Pre-Training.} We initialize the model with weights from ZipVoice~\citep{zhu2025zipvoice}, which is pre-trained on extensive monologue corpora. This stage establishes a robust foundation for speech-text alignment. Starting with this pre-trained model prevents the alignment failures observed when training directly on complex dialogue data.

\textbf{Stage 2: Dialogue Fine-Tuning.} The model is then fine-tuned on dialogue data to capture conversational dynamics. Specifically, the model adapts its speech-text alignment for multi-speaker contexts, learns to assign the correct timbre each speaker's turn, and generates natural turn-taking events.

\subsection{Speaker-Turn Embeddings}

To improve speaker turn-taking accuracy, ZipVoice-Dialog facilitates speaker disambiguation by incorporating learnable speaker-turn embeddings into the text features after the text encoder. Note that these embeddings differ from those used in speaker recognition, such as i-vectors~\citep{dehak2010front} or x-vectors~\citep{snyder2018x}. Instead, our approach employs two randomly initialized embeddings to denote the two speaker identities that are inferred from speaker-turn labels [S1] and [S2]. These two embeddings are optimized end-to-end alongside all other model components.

Specifically, for each text token $y_i$ in the interleaved text sequence, a corresponding speaker-turn embedding $e_{\text{speaker}(i)}$ is retrieved based on the binary speaker identity (i.e., speaker 1 or speaker 2). This speaker-turn embedding is then added to the text feature $\hat{y_i}$ before being expanded via average upsampling. The resulting text feature $\widetilde{y_i}$ for each token $y_i$ is therefore:
\begin{equation}
\widetilde{y_i} = \hat{y_i} + e_{\text{speaker}(i)}
\end{equation}

This mechanism allows the vector field estimator of ZipVoice-Dialog to accurately differentiate between speakers and assign the correct speaker voice to each turn.

\subsection{Input Format}

We describe the input format to enable ZipVoice-Dialog to handle the complexities of multi-speaker interactions.

(1) \textbf{Interleaved Text Input:} In dialogue datasets, speaker turns often overlap. To ensure the completeness of each speaker turn while maintaining the chronological order, we utilize a single chronologically interleaved text sequence~\citep{zhang2024covomix} as the text input. To construct this sequence, multi-turn utterances are first sorted by their start time. Adjacent utterances from the same speaker are then consolidated into a single turn, prefixed with a speaker identity token ([S1] or [S2]). Crucially, ZipVoice-Dialog models token and turn durations implicitly through the flow-matching objective, eliminating the need for pre-defined timestamps or an external token/turn duration predictor.

(2) \textbf{Speech Prompt with a Flexible Number of Speaker Turns:} During training, ZipVoice-Dialog adopts an in-filling strategy. A random-length prefix of the ground-truth dialogue, which may encompass multiple speaker turns, is considered as a speech condition. This formatting allows ZipVoice-Dialog to support flexible prompting during inference: users can provide a prompt speech with flexible number of speaker turns to guide the style and speaker identities of the generated dialogues.

\subsubsection{Extension for Stereo Dialogue generation}

For applications requiring clear speaker separation, such as immersive media or training full-duplex systems~\citep{defossez2024moshi}, generating stereo dialogue is highly desirable. To this end, we designed an extension of ZipVoice-Dialog, termed ZipVoice-Dialog-Stereo, which renders each speaker on a distinct channel. The methodology and corresponding experimental results for this extension are detailed in \autoref{subsec:stereo}.

\begin{figure*}[htbp]
    \centering
    \begin{subfigure}[b]{0.32\textwidth}
        \centering
        \includegraphics[width=\textwidth]{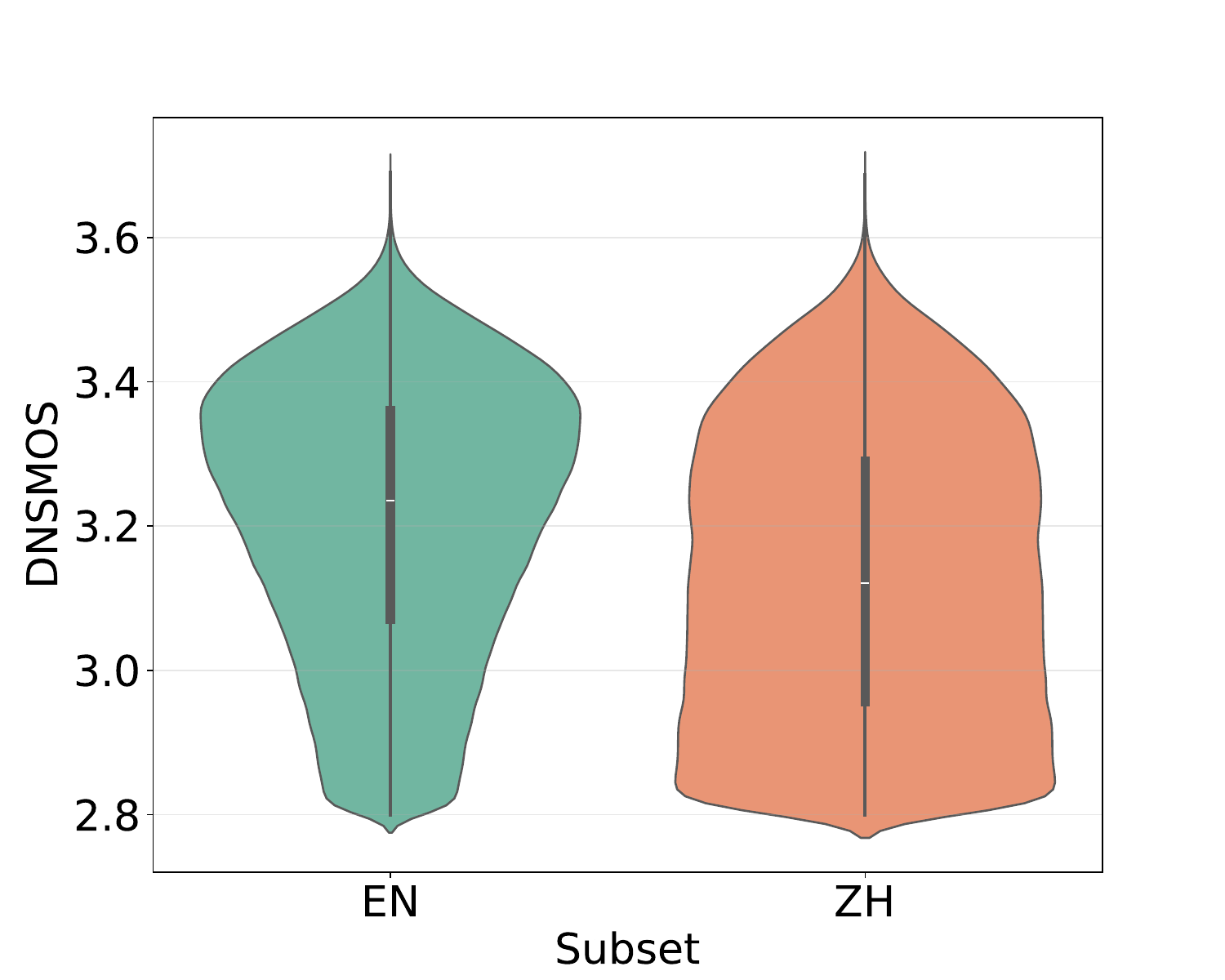}
        \caption{DNSMOS score}
        \label{fig:sub1}
    \end{subfigure}
    \begin{subfigure}[b]{0.32\textwidth}
        \centering
        \includegraphics[width=\textwidth]{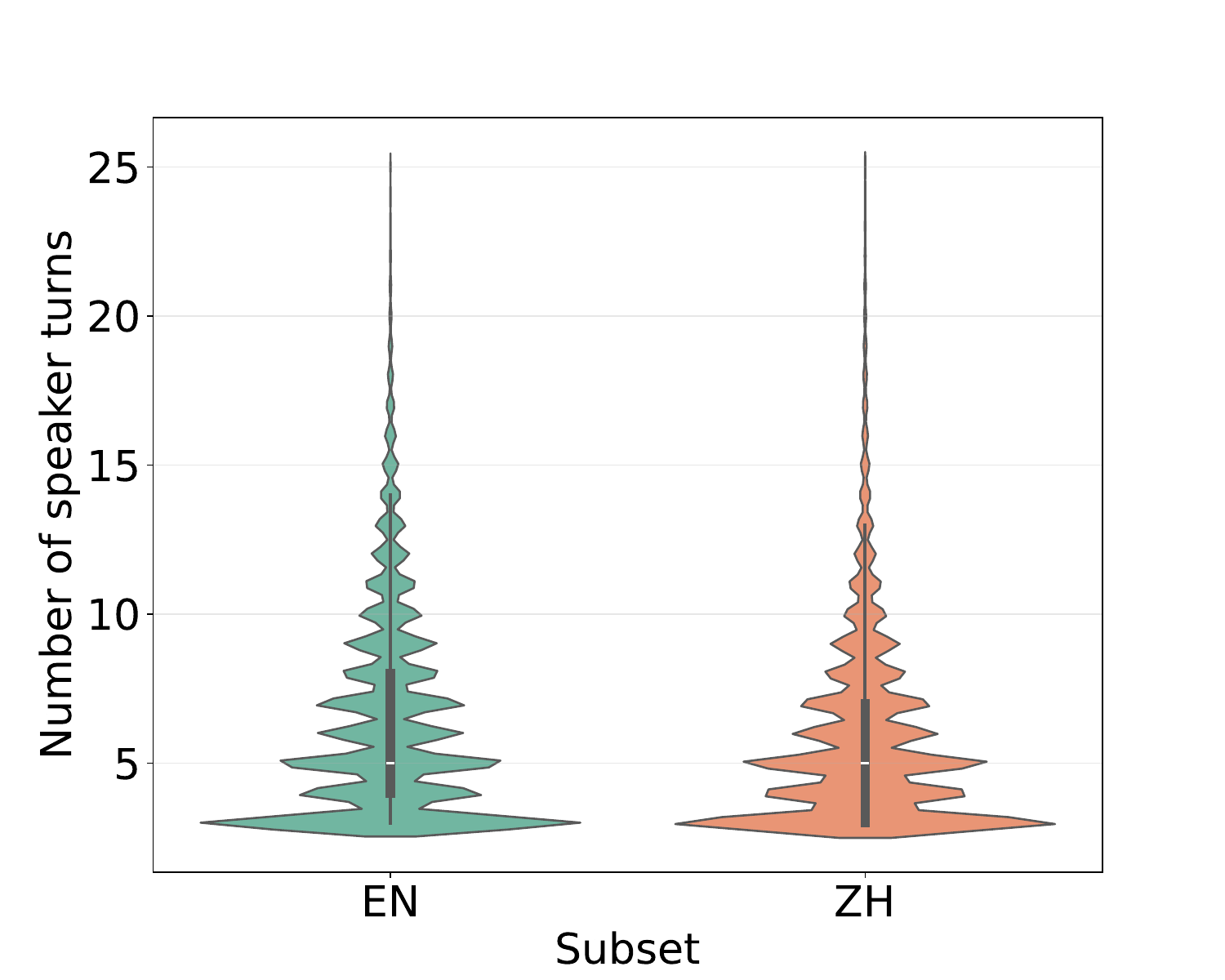}
        \caption{Number of speaker turns}
        \label{fig:sub2}
    \end{subfigure}
    \begin{subfigure}[b]{0.32\textwidth}
        \centering
        \includegraphics[width=\textwidth]{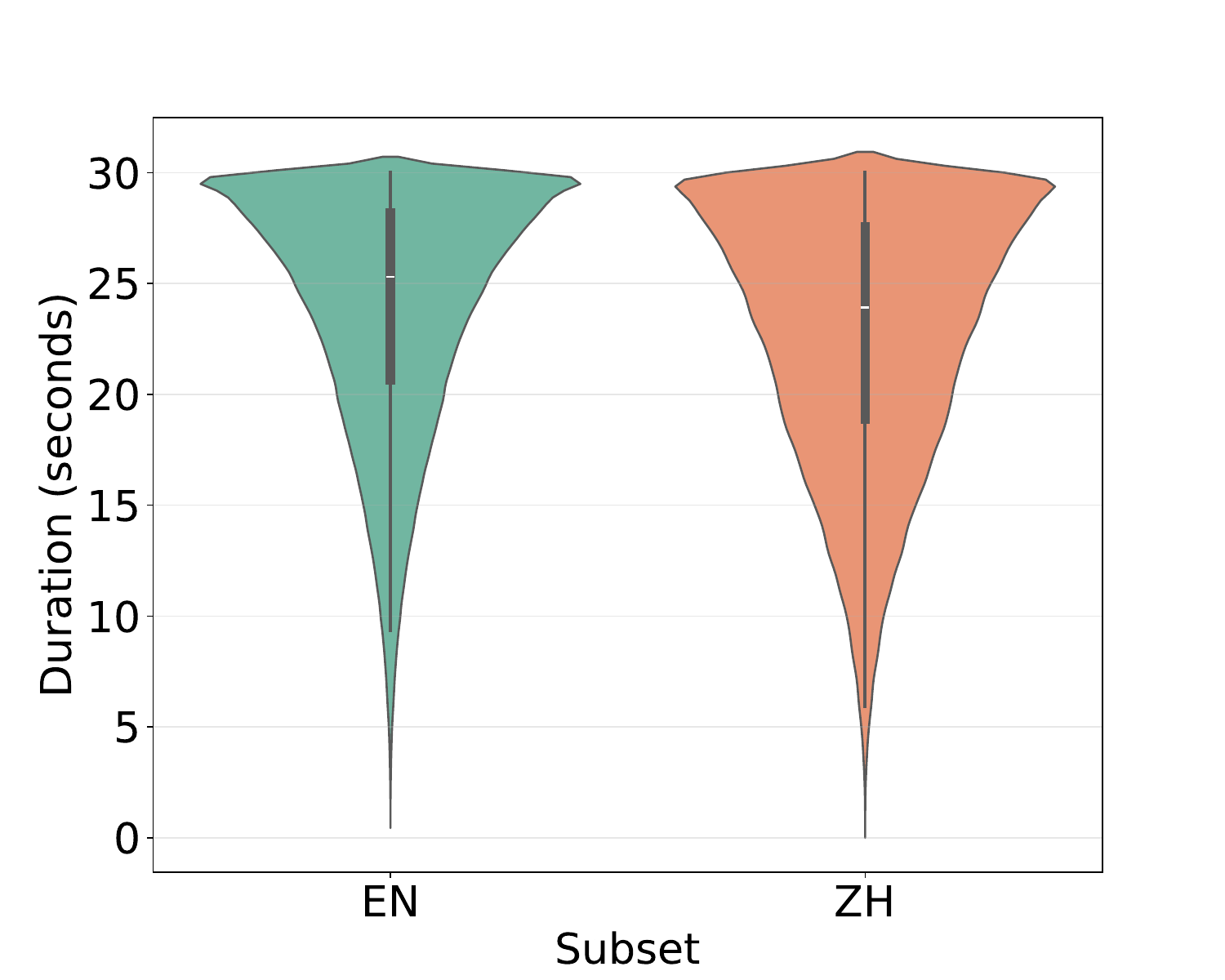}
        \caption{Duration per dialogue}
        \label{fig:sub3}
    \end{subfigure}
    \caption{Violin plots on English (EN) and Chinese (ZH) subsets of the OpenDialog dataset.}
    \label{fig:opendialog_statistic}
\end{figure*}

\section{OpenDialog Dataset}

The advancement of spoken dialogue systems is often hindered by a shortage of large, open-source datasets. To fill this gap, we introduce OpenDialog, a large-scale spoken dialogue corpus derived from real-world, in-the-wild speech data. This section details how the dataset was constructed.

\subsection{Mining In-the-Wild Spoken Dialogues}

The first major task was to mine spoken dialogue data from a large corpus of in-the-wild audios, which contains dialogue and non-dialogue content.

To isolate dialogues, we developed a multi-stage pipeline. First, we process the raw audio to identify and transcribe human speech. This begins with voice activity detection (VAD) to filter out silence and background noise, followed by speaker diarization to assign the resulting speech segments to distinct speakers. An automatic speech recognition (ASR) system then transcribes these segments into text, yielding a complete transcript with speaker labels. 
Finally, we leveraged a large language model as a classifier. The model assesses the semantic content and interactive nature of the speaker-attributed transcriptions to identify dialogues.

\subsection{Speaker-Attributed Transcription with WhisperD}

Once dialogue files were identified, the subsequent challenge was to obtain accurate speaker-attributed transcriptions.
We observed that the initial diarization-based speaker-attributed transcription is inaccurate in assigning speaker labels to short utterances, such as back-channels, which are frequently present in dialogues. To overcome this limitation, we use the WhisperD~\citep{darefsky2024parakeet} model, a fine-tuned Whisper~\citep{radford2023robust} model for speaker-attributed transcription of spoken dialogues.

For English data, we utilized the official open-source WhisperD model. For Chinese data, we trained a custom Chinese WhisperD model by fine-tuning Whisper on our in-house dialogue data. Since WhisperD is constrained by a 30-second input limit, we segmented longer recordings into compliant chunks. 
Long-form audios were first split based on silence via VAD, and any remaining segments longer than 30 seconds were further divided to meet the requirement.

\subsection{Rule-based Filtering}

The transcriptions obtained from the WhisperD ASR model are not perfect. To ensure transcription quality, we applied a set of rules to filter out data exhibiting abnormal transcription patterns, such as an abnormal number of speaker turns, unexpected language symbols, unusual word counts, excessive repetition, and improbable maximum word lengths.

Following this rule-based filtering, the remaining dialogue data was annotated with the DNSMOS P.835 OVRL score \citep{reddy2022dnsmos} and segments with a DNSMOS score less than 2.8 were removed.

Finally, we obtained a 6.8k-hour spoken dialogue dataset, comprising 1759 hours of Chinese data and 5074 hours of English data. The distributions of DNSMOS scores, number of speaker turns, and duration on English (EN) and Chinese (ZH) subsets are presented as violin plots in Figure~\ref{fig:opendialog_statistic}.

\section{Evaluation Benchmark} 
To evaluate the proposed model against existing baselines, we established a dedicated evaluation benchmark, including test sets from out-of-domain real-world spoken dialogue datasets and a suite of evaluation metrics covering various aspects.

\subsection{Test Sets}

We curated two test sets, test-zh (Chinese) and test-en (English), from two open-source real-world spontaneous spoken dialogue datasets \citep{yang2022open,zhou2025open}. test-zh includes 357 dialogues (2.23 hours), and test-en has 280 dialogues (1.84 hours). Each dialogue has a corresponding two-speaker-turn dialogue as the prompt. Derived from real-world speech, these sets capture authentic conversational styles. As they are unseen by the evaluated models, they serve as out-of-domain tests, enabling more challenging and practical assessments of spoken dialogue generation performance.

\subsection{Objective Evaluation Metrics}

We employed several objective metrics to facilitate fast and reproducible evaluation. These metrics are detailed as follows:

(1) \textbf{Intelligibility (WER):} We utilized the Word Error Rate (WER) calculated by comparing the transcription of the synthesized dialogue with the ground-truth text, irrespective of speaker identity. WhisperD~\citep{darefsky2024parakeet} was used to transcribe test-en, and Paraformer-zh~\citep{gao2022paraformer} was employed for test-zh. Speaker turn-taking symbols generated by WhisperD ([S1], [S2]) were ignored for this metric.

(2) \textbf{Speaker Turn-Taking Accuracy (cpWER):} To assess speaker turn-taking accuracy, i.e., whether the correct speaker voice is attributed to each utterance, we used concatenated minimum permutation word error rate (cpWER)~\citep{watanabe2020chime}. 
cpWER is computed by first concatenating utterances per speaker for reference and hypothesis files, respectively, then computing WERs between the reference and all possible speaker permutations of the hypothesis. The lowest WER among these permutations is cpWER. Speaker attribution errors lead to deletions for the correct speaker and insertions for the incorrect one. The gap between cpWER and standard WER thus measures speaker turn-taking accuracy, with a larger gap indicating more misattributed turns. We use WhisperD to obtain speaker-attributed transcriptions due to its high accuracy in identifying speaker turns. English dialogues under 30 seconds are evaluated (test-en (short)) to comply with Whisper's length constraint. Chinese dialogues are excluded due to Whisper's reduced performance on Chinese.

(3) \textbf{Speaker Similarity (cpSIM):} For speaker similarity assessment, we first employed a speaker diarization model~\citep{bredin2020pyannote} to segment generated dialogues into two speakers. Concatenated speech of each speaker was then separately fed into a WavLM-based~\citep{chen2022wavlm} ECAPA-TDNN model~\citep{desplanques2020ecapa} to extract speaker embeddings. Similarly, speaker embeddings of prompt speech were also extracted. We then computed the maximum speaker permutation cosine distance between speaker embeddings of generated and prompt speech. We refer to this metric as cpSIM, which stands for concatenated maximum permutation speaker similarity.

(4) \textbf{UTMOS:} UTMOS \citep{saeki2022utmos} is a neural network-based Mean Opinion Score (MOS) prediction model widely adopted to assess the quality of speech. Note that UTMOS scores for dialogue speech are lower than those for monologue speech, due to the metric's inherent bias toward English monologue.

(5) \textbf{Inference Speed (RTF):} To measure inference speed, we calculated the average real-time factor (RTF) on the test-en test set on GPUs.

\subsection{Subjective Evaluation Metrics}

As a complement to the objective evaluation metrics, two subjective evaluation metrics are used for subjective evaluation. 

(1) \textbf{Comparative Mean Opinion Scores (CMOS):} Evaluators were instructed to judge the relative quality of the dialogues in the range $[-3, 3]$, considering speaker turn-taking accuracy, coherence of speaker voices, fidelity to the input text condition, and naturalness.

(2) \textbf{Similarity Mean Opinion Scores (SMOS):} 
Evaluators were asked to rate the speaker similarity between the prompt and the generated dialogues in the range $[0, 5]$.

\begin{table*}[t]
    \centering
    \resizebox{\textwidth}{!}{
    \begin{tabular}{l *{8}{c}}
        \toprule
        \multirow{2}{*}{\textbf{Training Method}} & \multicolumn{3}{c}{\textbf{test-zh}} & \multicolumn{3}{c}{\textbf{test-en}} & \multicolumn{2}{c}{\textbf{test-en (short)}} \\
        \cmidrule(lr){2-4} \cmidrule(lr){5-7} \cmidrule(lr){8-9}
        & \textbf{cpSIM} $\uparrow$ & \textbf{WER} $\downarrow$ & \textbf{UTMOS} $\uparrow$ & \textbf{cpSIM} $\uparrow$ & \textbf{WER} $\downarrow$ & \textbf{UTMOS} $\uparrow$ & \textbf{WER} $\downarrow$ & \textbf{cpWER} $\downarrow$ \\
        \midrule
        w/ curriculum learning & \textbf{0.567} & \textbf{4.16} & \textbf{2.31} & \textbf{0.444} & \textbf{5.47} & \textbf{3.28} & \textbf{5.07} & \textbf{5.82} \\
        w/o curriculum learning & 0.424 & 84.19 & 1.78 & 0.247 & 116.10  & 1.87 & 116.65 & 116.31 \\
        \bottomrule
    \end{tabular}
    }
    \caption{Evaluation results of ZipVoice-Dialog models with and without curriculum learning.}
    \label{tab:curriculum}
\end{table*}

\begin{table}[htbp]
    \centering
    \resizebox{\columnwidth}{!}{
    \begin{tabular}{lcc}
        \toprule
        \multirow{2}{*}{\textbf{Method}} & \multicolumn{2}{c}{\textbf{test-en (short)}} \\
        \cmidrule(lr){2-3}
        & \textbf{WER} $\downarrow$ & \textbf{cpWER} $\downarrow$ \\
        \midrule
        Speaker turn-taking token | & 5.34 & 37.82 \\
        Speaker turn-taking tokens {[}S1{]} {[}S2{]} & 5.57 & 31.34 \\
        speaker-turn embedding & \textbf{5.07} & \textbf{5.82} \\
        \bottomrule
    \end{tabular}
    }
    \caption{Comparison of different speaker disambiguation methods.}
    \label{tab:speaker_disambiguation}
\end{table}

\section{Experimental Setup}

\subsection{Training Datasets} 

We trained ZipVoice-Dialog on two distinct datasets. The primary and larger dataset is the 6.8k-hour single-channel spoken dialogue dataset, OpenDialog. Complementing this, we utilized a smaller, two-channel in-house dataset annotated by human transcribers, comprising 736 hours of Chinese dialogue data and 84 hours of English dialogue data.

\subsection{Implementation Details}

ZipVoice-Dialog is trained by fine-tuning ZipVoice, a model pre-trained on 100k hours of monologue speech dataset~\citep{he2024emilia}, on single-channel dialogues for 60k updates with a total batch size of 4k seconds.
Inference was performed using 16 sampling steps with a Euler solver.

\section{Experimental Results}

In this section, we first verify the key methods in ZipVoice-Dialog model by training them on a small in-house dataset. Then, we examine the effectiveness of the constructed OpenDialog dataset by comparing models trained on different data. Finally, we compare ZipVoice-Dialog trained on full dataset against other SOTA spoken dialogue generation systems.

\subsection{Effectiveness of Curriculum Learning}

We attempted to train the ZipVoice-Dialog model from scratch using spoken dialogues. However, this approach consistently resulted in unintelligible speech. Specifically, while the generated audio sounded like human speech and could follow the style of the prompt speech, it failed to reflect the content of the input text.

As illustrated in Table~\ref{tab:curriculum}, the model trained without curriculum learning (i.e., without initialization from the monologue ZipVoice model) demonstrated reasonable results in terms of speaker similarity (cpSIM) and UTMOS. This indicates that the generated speech remained human-like and could mimic the prompt's characteristics. Nevertheless, the output speech remained largely unintelligible, as evidenced by an exceptionally high WER.

\subsection{Effectiveness of Speaker-Turn Embeddings}

We conducted experiments to evaluate different methods for speaker disambiguation within our architecture.
Table~\ref{tab:speaker_disambiguation} shows the impact of different speaker disambiguation methods. When we use a speaker turn-taking token "$|$" to separate speaker turns, the resulting speaker turn-taking accuracy was notably low (indicated by a significantly larger cpWER compared to WER). Switching to the use of two distinct speaker turn-taking tokens, "[S1]" and "[S2]", to precede their respective speaker turns, led to an improvement in speaker accuracy due to reduced ambiguity. However, this approach remained unsatisfactory. Crucially, the integration of speaker-turn embeddings led to a substantial reduction in cpWER, demonstrating a substantial improvement in speaker turn-taking accuracy.

\subsection{Impact of Different Training Data}

In this section, we analyze the impact of different training datasets on model performance, examining the effectiveness of the OpenDialog dataset.

\begin{table*}[t]
    \centering
    \resizebox{\textwidth}{!}{
    \begin{tabular}{l *{8}{c}}
        \toprule
        \multirow{2}{*}{\textbf{Training Data}} & \multicolumn{3}{c}{\textbf{test-zh}} & \multicolumn{3}{c}{\textbf{test-en}} & \multicolumn{2}{c}{\textbf{test-en (short)}} \\
        \cmidrule(lr){2-4} \cmidrule(lr){5-7} \cmidrule(lr){8-9}
        & \textbf{cpSIM} $\uparrow$ & \textbf{WER} $\downarrow$ & \textbf{UTMOS} $\uparrow$ & \textbf{cpSIM} $\uparrow$ & \textbf{WER} $\downarrow$ & \textbf{UTMOS} $\uparrow$ & \textbf{WER} $\downarrow$ & \textbf{cpWER} $\downarrow$ \\
        \midrule
        OpenDialog (6.8k) & 0.522 & \textbf{2.86} & 2.21 & 0.428 & 3.34 & 3.04 & \textbf{2.61} & 3.53 \\
        In-house dataset (0.8k) & \textbf{0.567} & 4.16 & \textbf{2.31} & \textbf{0.444} & 5.47 & \textbf{3.28} & 5.07 & 5.82 \\
        All (7.6k) & 0.556 & 3.17 & 2.25 & 0.437 & \textbf{3.25} & 3.07 & 2.79 & \textbf{3.27} \\
        \bottomrule
    \end{tabular}
    }
    \caption{Evaluation results of ZipVoice-Dialog models trained with different datasets.}
    \label{tab:training_data_comparison}

\end{table*}

As detailed in Table~\ref{tab:training_data_comparison}, the model trained exclusively on the larger OpenDialog dataset exhibits superior performance in terms of intelligibility (WER) compared to the model trained solely on the smaller in-house dataset. However, a slight degradation in speaker similarity (cpSIM) and UTMOS is observed when using OpenDialog alone. This is likely due to the higher speech quality and speaker annotation accuracy of our manually annotated in-house data. Moreover, combining both datasets during training yields a better balance across metrics. Crucially, models trained with either dataset demonstrate performance comparable to or exceeding existing baselines, suggesting that the performance of ZipVoice-Dialog is not highly sensitive to the size of training data. These experiments further indicate that our OpenDialog dataset alone is sufficient for training a high-performing spoken dialogue generation model.

\subsection{Comparison with Open-Sourced models}

\begin{table*}[t!]
    \centering
    \resizebox{\textwidth}{!}{
    \begin{tabular}{l c c *{8}{c}}
        \toprule
        \multirow{3}{*}{\textbf{Model}} & \multirow{3}{*}{\textbf{Params}} & \multirow{3}{*}{\textbf{RTF}$\downarrow$} & \multicolumn{3}{c}{\textbf{test-zh}} & \multicolumn{3}{c}{\textbf{test-en}} & \multicolumn{2}{c}{\textbf{test-en (short)}} \\
        \cmidrule(lr){4-6} \cmidrule(lr){7-9} \cmidrule(lr){10-11}
        & & & \textbf{cpSIM} $\uparrow$ & \textbf{WER} $\downarrow$ & \textbf{UTMOS} $\uparrow$ & \textbf{cpSIM} $\uparrow$ & \textbf{WER} $\downarrow$ & \textbf{UTMOS} $\uparrow$ & \textbf{WER} $\downarrow$ & \textbf{cpWER} $\downarrow$ \\
        \midrule
        Dia & 1.61B & 1.663 & - & - & - & 0.333 & 11.80 & 1.87 & 11.80 & 12.59 \\
        MoonCast & 2.67B & 0.953 & 0.463 & 15.85 & 1.78 & 0.356 & 23.62 & 2.37 & 8.41 & 16.53 \\
        ZipVoice-Dialog & 123M & \textbf{0.063} & \textbf{0.556} & \textbf{3.17} & \textbf{2.25} & \textbf{0.437} & \textbf{3.25} & \textbf{3.07} & \textbf{2.79} & \textbf{3.27} \\
        \bottomrule
    \end{tabular}
    }
    \caption{Objective performance comparison of different spoken dialogue generation models.}
    \label{tab:objective_metrics}
\end{table*}

We compared our proposed ZipVoice-Dialog model against two strong zero-shot spoken dialogue generation models: MoonCast~\citep{ju2025mooncast} and Dia~\citep{DiaRepo} where Dia is an open-source model conceptually similar to Parakeet~\citep{darefsky2024parakeet}. These two models represent two typical architectures. MoonCast employs a hybrid AR/NAR architecture. It first leverages a large language model as a text-to-semantic model to generate semantic speech tokens~\citep{wang2024maskgct}. This is followed by a flow-matching-based model for semantic-to-mel spectrogram reconstruction, and finally, a pre-trained vocoder for mel-to-waveform synthesis. In contrast, Dia is a purely AR model that directly predicts DAC audio tokens \citep{kumar2023high} and reconstructs the waveform from these tokens.

As presented in Table~\ref{tab:objective_metrics}, ZipVoice-Dialog consistently outperforms both MoonCast and Dia across all objective evaluation metrics. This superior performance is evident in objective measures such as inference speed (RTF), intelligibility (WER), speaker turn-taking accuracy (quantified by the difference between WER and cpWER), speaker similarity (cpSIM), and UTMOS.

Specifically, in terms of RTF, ZipVoice-Dialog achieves inference speeds over 15 times faster than that of the baselines. This notable efficiency stems from its considerably smaller model size and the inherent advantages of its NAR architecture.

Furthermore, regarding WER, Dia and MoonCast exhibit significantly worse performance, due to frequent word skipping and the generation of prolonged unintelligible segments, which are common instability issues in AR models. In contrast, ZipVoice-Dialog demonstrates greater stability, leading to its substantially lower WER.

\begin{table}[t!]
    \centering
    \resizebox{0.7\columnwidth}{!}{
    \begin{tabular}{l c c}
        \toprule
        \textbf{Model} & \textbf{CMOS} & \textbf{SMOS} \\
        \midrule
        MoonCast & -1.17 $\pm$ 0.12 & 2.35 $\pm$ 0.14 \\
        ZipVoice-Dialog & \textbf{0.00} & \textbf{3.86 $\pm$ 0.11} \\
        \bottomrule
    \end{tabular}
    }
    \caption{Subjective performance comparison of different spoken dialogue generation models.}
    \label{tab:subjective_metrics}

\end{table}

In terms of subjective quality, 10 Chinese native speakers are invited to evaluate the CMOS and SMOS on the test-zh subset. Each speaker evaluated 20 samples randomly sampled from this test set. As shown in Table~\ref{tab:subjective_metrics}, although MoonCast offers better expressiveness thanks to their larger parameters, its AR-based architecture results in various instability issues, considerably reducing the perceived subjective quality.

\section{Conclusion}

This paper introduces ZipVoice-Dialog, a NAR flow-matching model for zero-shot spoken dialogue generation. We proposed a curriculum learning strategy to address speech-text alignment challenges and incorporate speaker-turn embeddings to ensure accurate turn-taking. These two simple yet effective designs enable flow-matching-based architectures to achieve robust dialogue generation capabilities. Furthermore, we curated and released the first large-scale open-source spoken dialogue dataset. We also established a comprehensive evaluation benchmark. Experiments demonstrate the effectiveness of our proposed methods and dataset, showing that ZipVoice-Dialog achieves superior stability and efficiency in producing spoken dialogues compared to existing AR-based models.

\section*{Limitations}

One limitation lies in the model and data scale. While our prioritization of a compact architecture ensures high inference speed, it inevitably places a ceiling on expressiveness. Future exploration of larger models and datasets may yield more expressive dialogue generation capabilities. 
Another limitation is that subjective evaluations were restricted to to Chinese due to the availability of native speakers. Nonetheless, we validate our method through extensive, reproducible objective benchmarks across all languages. Moreover, we concentrate on two-speaker dialogue speech in this work, but the proposed methods are not constrained to two speakers and can generalize to multi-speaker dialogue.


\bibliography{bib}

\appendix

\section{Appendix}
\label{sec:appendix}

\subsection{ZipVoice-Dialog-Stereo: An Extension for Generating Stereo Dialogues}
\label{subsec:stereo}

We describe the methods that enables the stereo dialogue generation ability in the following sections.

\subsubsection{Inheriting Single-Channel Weights} 

To generate stereo dialogue, we adapt our model architecture by doubling the input and output feature dimensions to $2 * F$, with each half corresponding to a separate channel.

Our stereo model is initialized with weights from the pre-trained single-channel ZipVoice-Dialogue model for knowledge transfer. Most weights are directly transferable, but the input and output projection layers need resizing to fit the doubled feature dimension. These layers are initialized by duplicating the corresponding weights from the single-channel model for each of the two channels.

This initialization strategy minimizes disruption to the pre-trained weights, accelerating convergence and improving overall model performance.

\subsubsection{Single-Channel Dialogue Regularization} 

To maintain dialogue quality and prevent overfitting on the limited two-channel data, we introduce a regularization technique during stereo model fine-tuning.

Specifically, we retain the original single-channel input and output projection layers alongside the new stereo projection layers. The model is thus equipped with two parallel sets of projections, one dedicated to single-channel generation and the other to two-channel generation. To train this architecture, we alternate between batches of two-channel and single-channel dialogue speech. This approach alleviates catastrophic forgetfulness of knowledge acquired from the large single-channel dataset.

\begin{table*}[t]
    \centering
    \resizebox{\textwidth}{!}{
    \begin{tabular}{l *{8}{c}}
        \toprule
        \multirow{2}{*}{\textbf{Model}} & \multicolumn{3}{c}{\textbf{test-zh}} & \multicolumn{3}{c}{\textbf{test-en}} & \multicolumn{2}{c}{\textbf{test-en (short)}} \\
        \cmidrule(lr){2-4} \cmidrule(lr){5-7} \cmidrule(lr){8-9}
        & \textbf{cpSIM} $\uparrow$ & \textbf{WER} $\downarrow$ & \textbf{UTMOS} $\uparrow$ & \textbf{cpSIM} $\uparrow$ & \textbf{WER} $\downarrow$ & \textbf{UTMOS} $\uparrow$ & \textbf{WER} $\downarrow$ & \textbf{cpWER} $\downarrow$ \\
        \midrule
        ZipVoice-Dialog-Stereo & \textbf{0.474} & \textbf{2.909} & 2.12 & \textbf{0.321} & \textbf{4.67} & 2.99 & \textbf{3.66} & \textbf{4.93} \\
        w/o speaker exclusive loss & 0.468 & 3.663 & 2.05 & 0.314 & 5.10 & 2.71 & 4.70 & 5.95 \\
        w/o single-channel regularization & 0.461 & 3.027 & \textbf{2.18} & 0.317 & 5.56 & 3.00 & 4.33 & 6.09 \\
        w/o single-channel initialization & 0.457 & 3.887 & 2.11 & 0.319 & 5.89 & \textbf{3.01} & 5.01 & 6.59 \\
        \bottomrule
    \end{tabular}
    }
    \caption{Ablation experiments of ZipVoice-Dialog-Stereo.}
    \label{tab:stereo_ablation}
\end{table*}

\subsubsection{Speaker Exclusive Loss}

We observed that in ZipVoice-Dialog-Stereo, one of the two channels typically has degraded quality when two speakers talk simultaneously. This issue likely stems from the limited model capacity. Rather than scaling up the model at the cost of computational efficiency, we address this by introducing a speaker exclusive loss to penalize speech overlap.

Specifically, for timestep $t$, noisy speech feature $x_t$, and vector field estimator output $v_t(x_t)$, we first estimate clean speech features via an Euler step: $\hat{x}_1 = x_t + (1-t)v_t(x_t)$. These features are split into channel-specific features $f^c_{i,j}$ (where $c \in \{0,1\}$ is the channel index, $i$ is the frame index, and $j$ is the feature dimension).

Frame energy for channel $c$ at index $i$ is:
\begin{equation}
E_i^c = \frac{1}{D} \sum_{j=1}^{D} f_{i,j}^c
\end{equation}

To account for volume and background noise variations across samples, we use an adaptive energy threshold $\tau$ (instead of a fixed value), determined by the median frame energy of ground-truth speech:
\begin{equation}
\tau = \text{Quantile}\left( \{ E^c_i \}, q \right)
\end{equation}
where $\{ E^c_i \}$ includes all frame energies from both channels of ground-truth speech, and we set $q=0.5$ (median) under the assumption that approximately $50\%$ of two-channel speech frames are silent.

The speaker-exclusive loss penalizes frames where both channels' energies exceed $\tau$:
\begin{equation}
\mathcal{L}_{\text{SE}} = \frac{1}{T} \sum_{i=0}^{T-1} \mathbb{I}(E^0_i > \tau \land E^1_i > \tau) \cdot (E^0_i - \tau)(E^1_i - \tau)
\end{equation}
where $\mathbb{I}(\cdot)$ is the indicator function.

Total training loss combines this with the primary CFM loss:
\begin{equation}
\mathcal{L} = \mathcal{L}_{\text{CFM-TTS}} + \lambda \cdot \mathcal{L}_{\text{SE}}
\end{equation}
where $\lambda$ is set to 1 in this work.

Note that since this strategy penalizes speech overlap, it is not suitable to apply the model trained with speaker exclusive loss for applications requiring overlapped speech.

\subsubsection{Inference with Single-Channel Prompts}

Although ZipVoice-Dialog-Stereo is designed to use with stereo prompts, a common use case is use two single-channel prompts, one for each speaker. In such scenarios, a two-channel prompt must be constructed from these single-channel prompts, requiring a signal to be supplied for the inactive channel.

Pairing a monaural prompt with pure digital silence significantly degrades speech quality. This is because artificial silence creates an out-of-distribution input pattern, as the model was trained solely on stereo recordings with natural background noise in both channels.

To resolve this domain mismatch, we use a simple yet effective approach: instead of artificial silence, we concatenate the monaural prompt with pre-recorded real ambient noise for the inactive channel. This aligns with the training data, thus preserving the quality of generated stereo dialogue.

\subsection{Experimental Results of ZipVoice-Dialog-Stereo}

We conducted ablation experiments to evaluate the effectiveness of designs in ZipVoice-Dialog-Stereo. ZipVoice-Dialog-Stereo model is obtained by fine-tuning the single-channel ZipVoice-Dialog model for an additional 25k updates on our in-house two-channel dialogue dataset

For evaluation, the generated two-channel speech was mixed into a single-channel speech. And the performance was assessed using the same protocols as single-channel spoken dialogues.

As presented in Table~\ref{tab:stereo_ablation}, the overall performance of the stereo model is less competitive than its single-channel counterpart, partially due to the limited availability of two-channel spoken dialogue data. 
Despite this, our results clearly show that the proposed speaker exclusive loss, single-channel dialogue regularization, and the weight initialization strategy are all effective. These improvements are particularly pronounced in metrics such as intelligibility, speaker turn-taking accuracy, and speaker similarity.

\subsection{Subjective Evaluation Details}

We conducted CMOS and SMOS subjective evaluations. 10 Chinese native speakers are invited to evaluate the CMOS and SMOS, each speaker evaluated 20 random samples from the entire test set. Evaluators were informed in detail about the guidelines and scoring
criteria for the CMOS/SMOS test.

For the CMOS evaluation, the instructions are:

\begin{itemize}

\item Please pay attention to the following points during the evaluation:

\item Speaker Turn Accuracy: Whether the voice used in each sentence corresponds to the labeled speaker.

\item Speaker Consistency: Whether the tone and timbre of each speaker remain consistent throughout the entire dialogue.

\item Content Accuracy: Whether the audio content is consistent with the text. Inconsistencies include incorrect characters, extra characters, or missing characters.

\item Voice Naturalness: Whether the dialogue audio is natural, fluent, and clear.

\item Note: For the best results, please wear headphones and conduct the evaluation in a quiet environment.

\item The audio quality of Generated Audio 2 is higher than that of Generated Audio 1, resulting in a positive score, and vice versa for a negative score. Choice answer from the following ones: $3$ (Much better) $\rightarrow 2$ (Better) $\rightarrow 1$ (Slightly better) $\rightarrow 0$ (The same) $\rightarrow -1$ (Slightly worse) $\rightarrow -2$ (Worse) $\rightarrow -3$ (Much worse).
\end{itemize}

For the SMOS evaluation, the instructions are:

\begin{itemize}
    \item During the evaluation, please focus solely on the similarity in timbre (voice color) and prosody (rhythm and intonation) between the reference speech and the generated speech, ignoring differences in content, grammar, audio quality, and other factors.

    \item Note: For the best results, please wear headphones and conduct the evaluation in a quiet environment.

    \item The more similar the timbre and prosody of the generated audio are to the reference audio, the higher your score will be, and vice versa. Choice answer from the following ones:  $5 \rightarrow$ Excellent, Timbre and prosody are highly similar. $4 \rightarrow$ Very Good. $3 \rightarrow$ Good, Timbre and prosody are similar, but may have some subtle, perceptible differences. $2 \rightarrow$ Fair. $1 \rightarrow$ Poor, Timbre and prosody are similar in some aspects, but differences are easily noticeable.
\end{itemize}

\subsection{Ethics Statements}

ZipVoice-Dialog is strictly a research project. We acknowledge that spoken dialogue generation technology could potentially be misused for unauthorized voice cloning or the creation of deceptive audio content. Therefore, we emphasize that this model is intended solely for research purposes and should be deployed with robust safety guardrails and content authentication. Furthermore, we advocate for the implementation of watermarking and the development of detection models to identify AI-generated speech.

The training dataset restricts use to non-commercial research and educational purposes only. To ensure data integrity, we performed manual spot-checks on a random subset of the data, confirming that no individual's identity is uniquely identifiable and that the content is free of offensive material. We are committed to ongoing maintenance of the dataset to
address any potential risks in the future.

\end{document}